# Class Augmented Semi-Supervised Learning for Practical Clinical Analytics on Physiological Signals

**Arijit Ukil, Soma Bandyopadhyay, Chetanya Puri, Rituraj Singh, Arpan Pal**

Embedded Systems and Robotics
Research and Innovation
Tata Consultancy Services
Kolkata, India
*{arijit.ukil, soma.bandyopadhyay, chetanya.puri, singh.rituraj, arpan.pal}@tcs.com*

## Abstract

Computational analysis on physiological signals would provide immense impact for enabling automated clinical analytics. However, the class imbalance issue where negative or minority class instances are rare in number impairs the robustness of the practical solution. The key idea of our approach is intelligent augmentation of minority class examples to construct smooth, unbiased decision boundary for robust semi-supervised learning. This solves the practical class imbalance problem in anomaly detection task for computational clinical analytics using physiological signals. We choose two critical cardiac marker physiological signals: Heart sound or Phonocardiogram (PCG) and Electrocardiogram (ECG) to validate our claim of robust anomaly detection of clinical events under augmented class learning, where intelligent synthesis of minority class instances attempt to balance the class distribution. We perform extensive experiments on publicly available expert-labelled MIT-Physionet PCG and ECG datasets that establish high performance merit of the proposed scheme, and our scheme fittingly performs better than the state-of-the-art algorithms.

## 1    Introduction

Internet of Things (IoT) has the potential to shift the paradigm of the healthcare ecosystem through remote health monitoring, automating basic health screening, and exploiting Artificial Intelligence (AI)-based knowledge development to build a smart healthcare system [14]. It is often argued that many of the deadly diseases, including cardio-vascular diseases are preventable if timely action is taken and continuous monitoring is ensured. In fact, highest human life loss is reported due to cardio-vascular problems [1]. With the availability of low-cost sensors and computationally powerful smartphones, the development of automated cardio-vascular disease management ecosystem would be inexpensive but highly useful for the prevention of cardio-vascular disease related human life loss [2, 4, 16, 18]. Our aim is to develop robust clinical analytics, specifically anomaly detection solution using our novel class augmented semi-supervised learning such that highly accurate detection of cardiac abnormality can be identified from ECG and PCG signals. There are number of challenges involved in developing such effective clinical analytics system using physiological signals.

1. We find that in many real-life applications, mostly positive labeled training class instances form majority, whereas instances from another class is negligible [6]. We address the class-imbalance issue due to rarity of minority class examples. We construct novel intelligent minority class augmentation method exploiting *l*-diversity that generates diverse set of synthetic minority class examples and minimizes the classifier bias towards majority class.

2. ECG, PCG signals are often corrupted with various kinds of anomalies in terms of noise, motion artifacts. Identification and removal of corruptions in such signals are needed for accurate clinical decision making. We demonstrate an effective method of noise removal of such physiological signals, which consequently enhances the clinical utility of automated analytics.

## 2   Related Works and Our Contribution

Researchers have attempted to address the class-imbalance issue from different directions [3, 5]. SMOTE, an over-sampling method [3] is one of the most widely techniques where minority class augmentation is performed. SMOTE and similar techniques synthesize the minority class examples by forward prediction based generation and has high probability of redundancy in the generated examples. One Class Support Vector Machine (OC-SVM) is an established semi-supervised classification method to solve the class-imbalance problem [5, 7]. Recently, deep networks gained considerable attention and authors in [8] have proposed Convolution Neural Network (CNN) based deep learning architecture to solve the classification problem of sensor signals. Our novel intelligent class augmentation algorithm ensures diversity in the generated examples through feed-back controlled approach and avoids redundant example generation and effectively provide superior semi-supervised learning.

## 3   Our Proposed Class Augmented Learning

Let, training data $\mathcal{X} = (\mathbb{X}_+, \mathbb{X}_-)$, where $\mathbb{X}_+$ be the available known-labeled positive training set and $\mathbb{X}_-$ be the known-labeled negative training set, $\mathbb{X}_+ = \{\mathbb{x}_i^+\}_{i=1}^{\Pi}$, $\mathbb{X}_- = \{\mathbb{x}_i^-\}_{i=1}^{\pi}$, where $\Pi \gg \pi$; $\mathbb{x}_i^+, \mathbb{x}_i^- \in \mathbb{R}^d$. Formally we state the problem as: Let, training data $\mathcal{X} = \{(x_i, y_i)\}_{i=1}^M$, where $x_i \in \mathbb{R}^d$: the training instances, $y_i \in \mathcal{Y} = \{+1, -1\}$: class labels. We synthesize minority class examples $\mathbb{X}_{--} = \{\mathbb{x}_i^{--}\}_{i=1}^{\Pi}$ to balance the class instances. The synthesized class examples $\mathcal{X}^+ = (\mathbb{X}_+, \mathbb{X}_{--})$ are used to for semi-supervised learning. Let, testing data $T = \{(t_i, y_i)\}_{i=1}^N$, with distribution $\mathcal{T}$ and $f(x)$ be the function to construct the decision boundary. So, the primary objective is to find $f^*$ that minimizes the expected risk of prediction [9] of the proposed class augmented learning (instead of training instances, we learn from augmented training instances to tackle the class imbalance problem):

$$f^* = \underset{f}{\mathrm{argmax}}\, \mathbb{E}_{(x,y) \sim \mathcal{T}}\, err(y, f(x = \mathcal{X}^+))$$

### 3.1   Intelligent Synthesis of Minority Class Examples

In order to balance the training class instances, over-sampling methods are deployed to combat the class-imbalance problem such that number of training samples of the classes are same while constructing the classification function. Our intelligent synthesis method is diversity-optimized by feed-back controlled way (where synthetic examples are investigated on their diversity property) to minimize the redundancy nuisance. First, we generate a superset of class examples $\mathbb{X}_{---} = \{\mathrm{x}_i^{---}\}_{i=1}^{\Pi++}$ as $\mathbb{X}_- \to \mathbb{X}_{---}$ and subsequently generates $\mathbb{X}_{--} = \{\mathrm{x}_i^{--}\}_{i=1}^{\Pi}$ by our intelligent diversity-aware approach, $\mathbb{X}_{--} \subset \mathbb{X}_{---}$.

1. *Super Negative Class ($\mathbb{X}_{---}$) Generation:* Let, $\mathcal{S}$ distinct states be present in the given minority class instances $\mathbb{X}_- = \{\mathrm{x}_i^-\}_{i=1}^{\pi}$ and $\varphi_{ij}$ be the state transition probability from state $i$ to $j$. We deploy Markov model to generate the synthetic signals $\mathbb{X}_{---} = \{\mathrm{x}_i^{---}\}_{i=1}^{\Pi++}$ by combining the states within the transition probabilities $\varphi_{ij}$ [13].

2. *Redundancy-minimized Minority Class Augmentation:* The state-of-the art algorithms like SMOTE, directly generate synthetic examples from $\mathcal{X}$ without considering the redundancy in the generation process.
We synthesize the minority class examples by trading-off the significance and redundancy of the super negative class examples with the assumption that the generated augmented class examples $\mathbb{X}_{--}$ would be similar yet not redundant with $\mathbb{X}_-$, but distinct from $\mathbb{X}_+$. The algorithm is as follows:

  - Find significance score $\eta$ of each of the super negative class examples ($\mathbb{X}_{---}$): We compute the mutual information $\eta = \mathbb{I}(x = \mathbb{X}_{---}; y = \mathbb{X}_-) = \sum_{x \in \mathbb{X}_{---}} \sum_{y \in \mathbb{X}_-} p(x,y) \log_2 \frac{p(x,y)}{p(x)p(y)}$ for each of $\mathbb{X}_{---}$ with each of $\mathbb{X}_-$. The more is the score $\eta_i$ of $\mathrm{x}_i^{---}$, more is the significance of the generated synthetic example.

  - We order $\mathbb{X}_{---} = \{\mathrm{x}_i^{---}\}_{i=1}^{\Pi++}$ based on its $\eta_i$ score, $\mathbb{X}_{---} \xrightarrow{\eta} \mathbb{X}_{---}^{\eta}$.

  - Find $l$-diversity for each of $\mathbb{X}_{---}^{\eta}$: We form $l$-diverse group from $\mathbb{X}_{---}^{\eta}$ using $l$-diversity algorithm, where $l$-diversity grouping is defined as [10]: A group is $l$-diverse iff each of the '$l$' different group $\mathcal{L}$'s entropy $\geq \log_2 l$: $\sum_{s \in S} p(\mathcal{L}, s) \log_2 \frac{1}{p(\mathcal{L}, s)} \geq \log_2 l$, where $l = \lceil \Pi^{1/4} \rceil$, where $p(\mathcal{L}, s)$ is the fraction of diverse group where group elements with values equal to $'s'$ exists.

  - $\mathbb{X}_{--} = \{\mathrm{x}_i^{--}\}_{i=1}^{\Pi}$ is formed from the each of the $l$ groups such that total $\Pi$ number of elements (synthetic signals from the superset with $\Pi++$ number of signals) are chosen from $\mathbb{X}_{---}^{\eta}$.

We illustrate our approach in Figure 1.

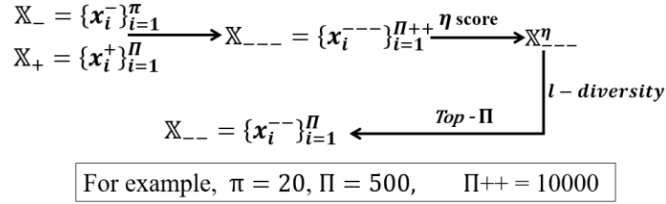

For example, $\pi = 20$, $\Pi = 500$, $\Pi++ = 10000$

Fig. 1. Illustration of proposed augmented class instance generation by maximizing significance score and minimizing redundancy

## 4    Experimental Results

We extensively experiment with expert-annotated publicly available PCG and ECG signals from MIT- Physionet database [15]. PCG dataset consists of more than 3500 instances with 1:5 minority, majority class example distribution, whereas ECG dataset consists of more than 8700 instances with 1:200 minority, majority class example distribution. Our semi-supervised learning based class augmentation method has two distinct improvements over the state-of-the-arts.

### 4.1 Validation of Augmented Class Generation

We show in Figure 2 that our proposed method of class augmentation for noisy physiological signal detection task outperforms the state-of-the-art methods: SMOTE [3] and classical One Class Support Vector Machine (OC-SVM) [5]. Performance comparison among the different algorithms are demonstrated by: Sensitivity (*se*), Specificity (*sp*), geometric mean of *se, sp* (*GM*). It is depicted in [17] that geometric mean of *se, sp* (*GM*) is a good indicator of a classifier performance under class imbalance scenario.

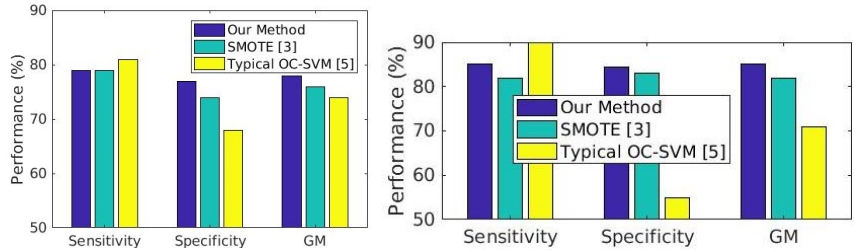

Fig. 2. Performance comparison of our algorithm with state-of-the-art [3, 5], (a) ECG, (b) PCG

### 4.2 Clinical Utility Enhancement

Computational clinical anomaly detection is performed to assess the cardiac health (whether a subject has cardiac disease or not). We show in Figure 3 that the clinical efficacy reported by our method is substantially higher when cardiac anomaly condition is detected from ECG and PCG signals. We compare our method with classical techniques (for ECG [11], and for PCG [12]) and related CNN-based deep learning method [8].

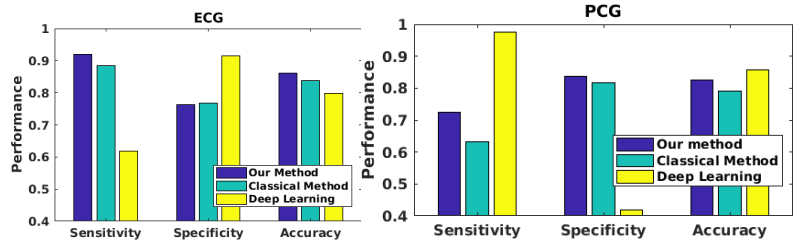

Fig. 3. Balanced performance of our proposed method, we note imbalance performance of CNN-based deep learning based method [8]

### 4.3 Observation

We observe that balanced performance of noise detection is better in PCG signals, which is probably due to the favorable class imbalance distribution. As a future scope of work, we shall attempt more robust techniques to ensure balanced performance in highly class imbalance scenario like the Physionet ECG datasets.

# References

[1] I. Silva, L. Joon Lee, and R. G. Mark. "Signal quality estimation with multichannel adaptive filtering in intensive care settings." IEEE Transactions on Biomedical Engineering, pp. 2476-2485, 2012.

[2] C. Liu, et al. "An open access database for the evaluation of heart sound algorithms," Physiological Measurement 37, no. 12, 2016.

[3] N. Chawla, et al. "SMOTE: synthetic minority over-sampling technique." Journal of artificial intelligence research, pp. 321-357, 2002.

[4] A. Ukil, S. Bandyopadhyay, C. Puri, and A. Pal, "Heart-trend: an affordable heart condition monitoring system exploiting morphological pattern," IEEE ICASSP, pp. 6260-6264, 2016.

[5] C.-C. Chang, C.-J. Lin, "Libsvm: a library for support vector machines," ACM transactions on intelligent systems and technology, 2011.

[6] G. Haixiang, L. Yijing, J. Shang, G. Mingyun, H. Yuanyue, and G. Bing, "Learning from class-imbalanced data," JESA, pp. 220 - 239, 2017.

[7] N. Shahid, I. Haider Naqvi, and S. Bin Qaisar, "One-class support vector machines: analysis of outlier detection for wireless sensor networks in harsh environments," JAIR, pp. 515-563, 2015.

[8] D. Palaz, M.M. Doss, R. Collobert, "Convolutional neural networks-based continuous speech recognition using raw speech signal," IEEE ICASSP, pp. 4295-4299, 2015.

[9] Q. Da, Y. Yu, and Z. Zhou, "Learning with augmented class by exploiting unlabeled data," AAAI, pp. 1760-1766, 2014.

[10] A. Machanavajjhala, D. Kifer, J. Gehrke, and M. Venkitasubramaniam, "L-diversity: Privacy beyond k-anonymity," ACM TKD, 2007.

[11] S. Datta, et al. "A Robust AF Classifier using Time and Frequency Features from Single Lead ECG Signal," IEEE CinC, 2017.

[12] C. Puri, A. Ukil, S. Bandyopadhyay, R. Singh, A. Pal, A. Mukherjee, and D. Mukherjee, "Classification of normal and abnormal heart sound recordings through robust feature selection." IEEE CinC, pp. 1125-1128, 2016.

[13] A. Ukil, S. Bandyopadhyay, C. Puri, R. Singh, and A. Pal, "On solving the class imbalance problem for clinical decision improvement using heart sound signals," IJCAI- LPCICD workshop, arXiv preprint arXiv:1707.09425, 2017.

[14] S. Bandyopadhyay, A. Ukil, C. Puri, R. Singh, A. Pal, CA Murthy, "Pattern Analysis in Physiological Pulsatile Signals: An Aid to Personalized Healthcare," IEEE EMBC, pp. 482-485, 2018.

[15] https://physionet.org/pn3/challenge/

[16] A. Ukil, and U. Kumar Roy, "Smart cardiac health management in IoT through heart sound signal analytics and robust noise filtering," IEEE PIMRC, 2017.

[17] X. Guo, Y. Yin, C. Dong, G. Yang, and G. Zhou, "On the Class Imbalance Problem," International Conference on Natural Computation, pp. 192 - 201, 2008.

[18] A. Ukil, S. Bandyopadhyay, C. Puri, R. Singh, A. Pal, A. Mukherjee, "Heartmate: automated integrated anomaly analysis for effective remote cardiac health management," IEEE ICASSP, pp.6578-6579, 2017.